\documentstyle[twoside]{article}

\catcode`\@=11
\long\def\@makefntext#1{
\protect\noindent \hbox to 3.2pt {\hskip-.9pt

$^{{\eightrm\@thefnmark}}$\hfil}#1\hfill}		

\def\@makefnmark{\hbox to 0pt{$^{\@thefnmark}$\hss}}	

\def\ps@myheadings{\let\@mkboth\@gobbletwo
\def\@oddhead{\hbox{}
\rightmark\hfil\eightrm\thepage}

\def\@oddfoot{}\def\@evenhead{\eightrm\thepage\hfil
\leftmark\hbox{}}\def\@evenfoot{}
\def\sectionmark##1{}\def\subsectionmark##1{}}

\oddsidemargin=\evensidemargin
\newcounter{sectionc}\newcounter{subsectionc}\newcounter{subsubsectionc}
\renewcommand{\section}[1] {\vspace{12pt}\addtocounter{sectionc}{1}

\setcounter{subsectionc}{0}\setcounter{subsubsectionc}{0}\noindent

	{\tenbf\thesectionc. #1}\par\vspace{5pt}}
\renewcommand{\subsection}[1]
{\vspace{12pt}\addtocounter{subsectionc}{1}

	\setcounter{subsubsectionc}{0}\noindent

	{\bf\thesectionc.\thesubsectionc. {\kern1pt \bfit
#1}}\par\vspace{5pt}}
\renewcommand{\subsubsection}[1]
{\vspace{12pt}\addtocounter{subsubsectionc}{1}
	\noindent{\tenrm\thesectionc.\thesubsectionc.\thesubsubsectionc.
	{\kern1pt \tenit #1}}\par\vspace{5pt}}
\newcommand{\nonumsection}[1] {\vspace{12pt}\noindent{\tenbf #1}
	\par\vspace{5pt}}

\newcounter{appendixc}
\newcounter{subappendixc}[appendixc]
\newcounter{subsubappendixc}[subappendixc]
\renewcommand{\thesubappendixc}{\Alph{appendixc}.\arabic{subappendixc}}
\renewcommand{\thesubsubappendixc}
	{\Alph{appendixc}.\arabic{subappendixc}.\arabic{subsubappendixc}}

\renewcommand{\appendix}[1] {\vspace{12pt}
        \refstepcounter{appendixc}
        \setcounter{figure}{0}
        \setcounter{table}{0}
        \setcounter{lemma}{0}
        \setcounter{theorem}{0}
        \setcounter{corollary}{0}
        \setcounter{definition}{0}
        \setcounter{equation}{0}
        \renewcommand{\thefigure}{\Alph{appendixc}.\arabic{figure}}
        \renewcommand{\thetable}{\Alph{appendixc}.\arabic{table}}
        \renewcommand{\theappendixc}{\Alph{appendixc}}
        \renewcommand{\thelemma}{\Alph{appendixc}.\arabic{lemma}}
        \renewcommand{\thetheorem}{\Alph{appendixc}.\arabic{theorem}}

\renewcommand{\thedefinition}{\Alph{appendixc}.\arabic{definition}}

\renewcommand{\thecorollary}{\Alph{appendixc}.\arabic{corollary}}

\renewcommand{\theequation}{\Alph{appendixc}.\arabic{equation}}
        \noindent{\tenbf Appendix \theappendixc #1}\par\vspace{5pt}}
\newcommand{\subappendix}[1] {\vspace{12pt}
        \refstepcounter{subappendixc}
        \noindent{\bf Appendix \thesubappendixc. {\kern1pt \bfit #1}}
	\par\vspace{5pt}}
\newcommand{\subsubappendix}[1] {\vspace{12pt}
        \refstepcounter{subsubappendixc}
        \noindent{\rm Appendix \thesubsubappendixc. {\kern1pt \tenit
#1}}
	\par\vspace{5pt}}

\newcommand{\textlineskip}{\baselineskip=13pt}
\newcommand{\smalllineskip}{\baselineskip=10pt}

\def\eightcirc{
\begin{picture}(0,0)
\put(4.4,1.8){\circle{6.5}}
\end{picture}}
\def\eightcopyright{\eightcirc\kern2.7pt\hbox{\eightrm c}}

\def\abstracts#1#2#3{{
	\centering{\begin{minipage}{4.5in}\baselineskip=10pt\footnotesize
	\parindent=0pt #1\par

	\parindent=15pt #2\par
	\parindent=15pt #3
	\end{minipage}}\par}}

\newcommand{\bibit}{\nineit}

\renewenvironment{thebibliography}[1]
	{\frenchspacing
	 \ninerm\baselineskip=11pt
	 \begin{list}{\arabic{enumi}.}
	{\usecounter{enumi}\setlength{\parsep}{0pt}
	 \setlength{\leftmargin 12.7pt}{\rightmargin 0pt} 
	 \setlength{\itemsep}{0pt} \settowidth
	{\labelwidth}{#1.}\sloppy}}{\end{list}}

\newcounter{itemlistc}
\newcounter{romanlistc}
\newcounter{alphlistc}
\newcounter{arabiclistc}

\newcommand{\fcaption}[1]{
        \refstepcounter{figure}
        \setbox\@tempboxa = \hbox{\footnotesize Fig.~\thefigure. #1}
        \ifdim \wd\@tempboxa > 5in
           {\begin{center}
        \parbox{5in}{\footnotesize\smalllineskip Fig.~\thefigure. #1}
            \end{center}}
        \else
             {\begin{center}
             {\footnotesize Fig.~\thefigure. #1}
              \end{center}}
        \fi}

\newcommand{\tcaption}[1]{
        \refstepcounter{table}
        \setbox\@tempboxa = \hbox{\footnotesize Table~\thetable. #1}
        \ifdim \wd\@tempboxa > 5in
           {\begin{center}
        \parbox{5in}{\footnotesize\smalllineskip Table~\thetable. #1}
            \end{center}}
        \else
             {\begin{center}
             {\footnotesize Table~\thetable. #1}
              \end{center}}
        \fi}

\def\@citex[#1]#2{\if@filesw\immediate\write\@auxout
	{\string\citation{#2}}\fi
\def\@citea{}\@cite{\@for\@citeb:=#2\do
	{\@citea\def\@citea{,}\@ifundefined
	{b@\@citeb}{{\bf ?}\@warning
	{Citation `\@citeb' on page \thepage \space undefined}}
	{\csname b@\@citeb\endcsname}}}{#1}}

\newif\if@cghi
\def\cite{\@cghitrue\@ifnextchar [{\@tempswatrue
	\@citex}{\@tempswafalse\@citex[]}}
\def\citelow{\@cghifalse\@ifnextchar [{\@tempswatrue
	\@citex}{\@tempswafalse\@citex[]}}
\def\@cite#1#2{{$\null^{#1}$\if@tempswa\typeout
	{IJCGA warning: optional citation argument

	ignored: `#2'} \fi}}

\def\pmb#1{\setbox0=\hbox{#1}
	\kern-.025em\copy0\kern-\wd0
	\kern.05em\copy0\kern-\wd0
	\kern-.025em\raise.0433em\box0}

\def\fnt#1#2{\footnotetext{\kern-.3em
	{$^{\mbox{\scriptsize #1}}$}{#2}}}

\def\fpage#1{\begingroup
\voffset=.3in
\thispagestyle{empty}\begin{table}[b]\centerline{\footnotesize #1}
	\end{table}\endgroup}

\font\tenrm=cmr10
\font\tenit=cmti10

\font\tenbf=cmbx10
\font\bfit=cmbxti10 at 10pt
\font\ninerm=cmr9
\font\nineit=cmti9

\font\eightrm=cmr8

\def\qed{\hbox{${\vcenter{\vbox{			
   \hrule height 0.4pt\hbox{\vrule width 0.4pt height 6pt
   \kern5pt\vrule width 0.4pt}\hrule height 0.4pt}}}$}}

\begin{document}


\normalsize\textlineskip
\thispagestyle{empty}
\setcounter{page}{1}

\vspace{-5.4in}
\rightline{CERN-TH/96-229}
\rightline{hep-th/9608171}

\vspace*{0.88truein}

\fpage{1}
\centerline{\bf CONDENSATION OF p-BRANES AND}
\vspace*{0.035truein}
\centerline{\bf GENERALIZED HIGGS/CONFINEMENT DUALITY\footnote{
To appear in the proceedings of the 1996 summer Telluride workshop.}}
\vspace*{0.37truein}
\centerline{\footnotesize FERNANDO QUEVEDO}
\vspace*{0.015truein}
\centerline{\footnotesize\it Theory Division, CERN}
\baselineskip=10pt
\centerline{\footnotesize\it CH-1211 Geneva 23, Switzerland}
\vspace*{10pt}
\centerline{\footnotesize CARLO A. TRUGENBERGER}
\vspace*{0.015truein}
\centerline{\footnotesize\it Department of Theoretical Physics,
University
of Geneva}
\baselineskip=10pt
\centerline{\footnotesize\it CH-1211 Geneva 4, Switzerland}
\vspace*{0.225truein}

\vspace*{0.21truein}
\abstracts{We review our recent work on the low-energy actions and
the realizations of strong-weak coupling dualities in
non-perturbative
phases of compact antisymmetric tensor field theories due to
p-brane condensation. As examples we derive and discuss the confining
string and confining membrane actions obtained from compact vector
and tensor theories in 4D. We also mention the relevance of
our results for the description of the Hagedorn phase transition of
finite temperature  strings.}{}{}

\textlineskip			
\vspace*{12pt}			

\vspace*{-0.5pt}
\noindent
Dualities play an ever increasing role in modern high-energy
physics.$^1$
Accordingly, both field theory and string dualities are the subject
of
current intensive investigations.

Field theory dualities can be roughly divided into Abelian and
non-Abelian
dualities, the former being much better understood since they can be
derived
from an action principle. The paramount examples of this {\it
strong-weak
coupling Abelian dualities} are given by a massless scalar field
theory in
2D and by 4D sourceless QED.

The natural generalization of these two examples is given by generic
antisymmetric tensor field theories$^{2,3}$ with (Euclidean) action
\begin{equation}
S=\int {1\over e^2} \ d\phi _{h-1}\wedge * d\phi _{h-1}\ ,
\label{a}
\end{equation}
where $\phi _{h-1}$ is an $(h-1)$-form in $D=d+1$ space-time
dimensions
and $e^2$ is a dimensionless coupling constant. The $U(1)$ gauge
invariance
under transformations $\phi _{h-1}\to \phi _{h-1} + d\lambda _{h-2}$
implies that (\ref{a}) describes $\left( {d-1\atop h-1} \right)$
massless
degrees of freedom.

The tensor $\phi _{h-1}$ couples naturally to a closed
$(h-2)$-dimensional
extended object: in modern parlance a $(h-2)$-brane:
\begin{eqnarray}
S & \to & S+i \tilde q \int \phi _{h-1} \wedge * \tilde j_{h-1} \ ,\\
\tilde j_{h-1}^{\mu _1 \dots \mu _{h-1}} & = &
\int \delta ^{d+1} \left( x- \tilde y (\tilde \sigma ) \right) \
d\tilde
y^{\mu _1} \wedge \dots \wedge d\tilde y^{\mu _{h-1}} \ ,
\label{b}
\end{eqnarray}
where $\tilde q$ is a coupling of canonical dimension $(2h-d-1)/2$
and
$\tilde y (\tilde \sigma )$ parametrizes the closed (or infinitely
extended)
world-hypersurface of the $(h-2)$-brane. Accordingly, (\ref{a}) can
be viewed
as an Abelian gauge theory for closed $(h-2)$-branes.

The model dual to (\ref{a}) is formulated in terms of a $(d-h)$-form
$\tilde \phi _{d-h}$ defined by $d\tilde \phi _{d-h} = *d\phi _{h-1}$
and
has the action
\begin{equation}
\tilde S=\int {e^2\over 4} \ d\tilde \phi _{d-h} \wedge *d\tilde
\phi_
{d-h} \ .
\label{d}
\end{equation}
Note that the coupling constant is reversed so that a strong-coupling
regime
of (\ref{a}) corresponds to a weak-coupling regime of (\ref{d}) and
viceversa. This model can be viewed as a gauge theory for closed (or
infinitely extended) $(d-h-1)$-branes:
\begin{eqnarray}
S & \to & S+iq \int \tilde \phi _{d-h} \wedge * j_{d-h} \ ,\\
j_{d-h}^{\mu _1 \dots \mu _{d-h}} & = & \int \delta ^{d+1} \left(
x-y(\sigma ) \right) \ dy^{\mu _1} \wedge \dots \wedge dy^{\mu
_{d-h}} \ ,
\label{e}
\end{eqnarray}
where the coupling $q$ has dimension $(d-2h+1)/2$.

Higher-rank antisymmetric tensors appear naturally in supersymmetric
field
theories and play an ubiquitous role in the field theory low-energy
limits
of various string theories.$^4$
The above duality transformation plays a key
role in the realization of strong-weak coupling
dualities among string theories.$^5$

The relation between (\ref{a}) and (\ref{d}) is the simplest example
of
the realization of an Abelian duality in field theory. In the
following
we wish to discuss a simple, yet highly non-trivial generalization
which is still tractable in general. This is obtained by considering
{\it compact} antisymmetric field theories, which are the
generalizations
of Polyakov's compact QED$^6$ to higher-rank tensor theories.$^7$

In compact antisymmetric field theories $p$-branes appear also as
{\it topological defects} of the original theory. For the model
(\ref{a})
these can be viewed e.g. as closed $(d-h)$-dimensional singularities
excluded from the definition domain of the (Euclidean) model. This is
then
considered as a low-energy effective field theory valid only
`outside'
these singularities, which represent the world-hypersurfaces of
$(d-h-1)$-branes.

The presence of these singularities induces non-trivial homology
cycles
for the antisymmetric tensor fields. Let us cut one of these
singularities
with an $(h+1)$-dimensional hyperplane $\Sigma _{h+1}$ and let us
choose a
sphere $S_h$ on $\Sigma _{h+1}$  around one of the two intersection
points
with the $(d-h)$-dimensional singularity. The topological quantum
number is
then given by $\int _{S_h} d\phi _{h-1}$. Note that in this
formulation
instantons appear as $(-1)$-branes.

As always for effective field theories, a proper definition of the
model
requires an ultraviolet cutoff. For vector theories one can obtain
the
compact $U(1)$ group by spontaneous symmetry breaking of a compact
non-Abelian group. In this case the cutoff is provided by the mass of
the
broken gauge fields. For higher-rank tensors this interpetation is no
more
possible. Keeping in mind that antisymmetric tensors appear in the
low-energy
limit of string theories one could think of
the cutoff as provided by massive string modes. In such models,
however,
the low-energy theory contains typically additional massless fields,
as
the graviton and the dilaton. The gravitational sector, in
particular,
shields the singularities by event horizons, leading to finite masses
of
solitonic $p$-branes already in the low-energy theory.$^8$ In the
following
we shall not include the graviton and the dilaton but deal rather
only
with compact antisymmetric tensors. When needed one can always resort
to a
lattice regularization, a procedure that works for any rank of the
tensors.

We shall consider both the original tensor $\phi _{h-1}$ and its dual
$\tilde \phi _{d-h}$ to be compact, so that we have two types of
topological
defects: $(d-h-1)$-branes for $\phi _{h-1}$ and $(h-2)$-branes for
$\tilde \phi _{d-h}$. In this case the dual actions (\ref{a}) and
(\ref{d})
have to be modified in order to take into account explicitly the
topological
defects. The new pair of dual actions is given by
\begin{eqnarray}
S &=& \int {1\over e^2} \ \left( d\phi _{h-1} -qV_h \right) \wedge *
\left( d\phi _{h-1} -qV_h \right) + i\tilde q \ \phi _{h-1} \wedge *
\tilde j_{h-1} \ ,\\
\tilde S &=& \int {e^2\over 4} \ \left(d\tilde \phi _{d-h} - \tilde q
\tilde V_{d+1-h} \right) \wedge * \left( d\tilde \phi _{d-h} - \tilde
q
\tilde V_{d+1-h} \right) +iq \ \tilde \phi _{d-h} \wedge * j_{d-h} ,
\label{g}
\end{eqnarray}
where the (singular) forms $V$ and $j$ describing the topological
defects
are defined as follows:
\begin{eqnarray}
*V_h & = & T_{d-h+1}\ , \qquad \qquad \qquad
\qquad \ \ j_{d-h}=\delta T_{d-h+1}\ ,\\
    {}*\tilde {V}_{d+1-h} & = & -(-1)^{h(d+1-h)} \ \tilde T_h \
,\qquad \qquad
\tilde j_{h-1} = \delta \tilde T_h \ .
\label{i}
\end{eqnarray}
Here
\begin{eqnarray}
T_{d-h+1}^{\mu _1\dots \mu _{d-h+1}} & = & \int \delta ^{d+1} \left(
x-y(\sigma ) \right) \ dy^{\mu _1} \wedge \dots \wedge dy^{\mu
_{d-h+1}} \ ,\\
\tilde T_h^{\mu _1\dots \mu _h} & = & \int \delta ^{d+1} \left(
x-\tilde y (\tilde \sigma ) \right) \ d\tilde y^{\mu _1} \wedge \dots
\wedge
d\tilde y^{\mu _h} \ .
\label{m}
\end{eqnarray}
Here $T_{d-h+1}$ and $\tilde T_h$ are the volume forms for open
hypervolumes
$y(\sigma )$ and $\tilde y(\tilde \sigma )$ which describe the
generalization
to higher-dimensional topological defects of the familiar Dirac
string
connecting a monopole-antimonopole pair. The boundaries of these
hypervolumes, described by the (tensor) currents $j_{d-h}$ and
$\tilde j_
{h-1}$, are the world-hypersurfaces of the topological defects. Note
that
these are now dynamical objects which have to be traced over in the
partition
functions corresponding to the above actions. The couplings $q$ and
$\tilde q$
represent the `charge' units of the $(d-h-1)$-branes and
$(h-2)$-branes
respectively: they are the generalization of the familiar magnetic
and electric
charge units $g$ and $e$. Establishing duality of the above pair of
actions
requires a generalized Dirac quantization condition
\begin{equation}
q\tilde q = 2\pi p\ , \qquad p\in Z \ .
\label{o}
\end{equation}
The solitonic $(d-h-1)$-branes of the compact
version of (\ref{a}) couple as `Noether currents' to the dual tensor
field $\tilde \phi _{d-h}$, as in (\ref{e}), and viceversa.

Topological defects can condense, leading to drastic modifications of
the infrared behaviour of the original theory.$^6$ When studying such
effects
two separated questions have to be answered. The first regards the
condensation dynamics and can be addressed best with a lattice
regularization:
the condensation mechanism is rather well established$^{6,7,9}$
for $h=d$ ($(-1)$-branes = instantons) and
for $h=d-1$ ($0$-branes = monopoles), although there
are also partial results on the condensation of
higher-dimensional branes.$^{10}$
The second question, which we would like to address here, regards the
nature of the new phase with a continuous distribution of topological
defects and the low-energy effective action describing it.

This question has been addressed nearly twenty years ago by Julia and
Toulouse$^{11}$ in the context of ordered solid state media. In a
nutshell,
their idea is that the condensation of topological defects gives rise
to new low-energy (hydrodynamical) modes representing the
long-wavelength
fluctuations about the homogeneous distribution of topological
defects.
While in the framework of ordered solid state media this idea did not
bring very far due to non-linearities and the need of introducing
dissipation terms, it is perfectly apt to study compact antisymmetric
tensor theories, where it naturally leads also to the effective
action for
the new phase.$^{12}$

Let us concentrate first on the original model (\ref{a}). In our
framework
the idea of Julia and Toulouse can be implemented simply by promoting
the singular form $V_h$ in (\ref{i}) to a new tensor field $\omega
_h$.
Since the conserved (tensor) current describing the fluctuations
about
the homogenoeus distribution of topological defects is
$j_{d-h} \propto *d\omega _h$, the new degrees of freedom are
associated
with the {\it gauge invariant part} of $\omega _h$.

Condensation of $(d-h-1)$-branes generates a new scale $\Lambda $.
This
can be taken as $\Lambda \propto \rho ^{1/h+1}$, where $\rho $ is the
average density of intersection points of the $(d-h)$-dimensional
world-hypersurfaces of the condensed branes with any
$(h+1)$-dimensional
hyperplane in the Euclidean space-time.

The three requirements on the effective action for the dense phase
are gauge
invariance under transformations $\omega _h \to \omega _h +d\psi
_{h-1}$,
relativistic invariance and recovering the original model in the
limit
$\Lambda \to 0$. Up to two derivatives in the new field we are thus
led to
the action
\begin{equation}
S_{d-h-1}=\int {1\over \Lambda ^2} \ d\omega _h \wedge *d\omega _h
+{1\over e^2}\left( \omega _h-d\phi _{h-1} \right) \wedge * \left(
\omega _h -d\phi _{h-1} \right) \ ,
\label{p}
\end{equation}
where gauge invariance is realized by accompanying transformations
$\omega _h \to \omega _h +d\psi _{h-1}$ with transformations
$\phi _{h-1} \to \phi _{h-1}+ \psi _{h-1}$.

This gauge invariance must be gauge fixed. As always for Abelian
theories,
this means that one can drop the integration over $\phi _{h-1}$ after
reabsorbing $d\phi _{h-1}$ into $\omega _h$, so that the action
describes
$\left( {d\atop h} \right) $ physical degrees of freedom of {\it
mass}
$m=\Lambda /e$. This mechanism, which we called the Julia-Toulouse
mechanism, is the exact opposite of the familiar Higgs-St\"uckelberg
mechanism: the new degrees of freedom, generated by the condensation
of
topological defects, `eat' the original degrees of freedom thereby
acquiring a mass.

In order to establish the nature of this phase we note that the
original
coupling (\ref{b}) to closed $(h-2)$-branes has to be promoted to a
new
coupling
\begin{equation}
S_{d-h-1} \to S_{d-h-1} + i\tilde q \int \omega _h \wedge * \tilde
T_h \ ,
\label{q}
\end{equation}
where $\tilde T_h$ describes an open hypervolume bounded by the
world-hypersurface of the original $(h-2)$-brane. Since $\omega _h$
is
a massive field, the induced action for $\tilde T_h$ reduces in the
infrared to
\begin{equation}
S_{\rm ind} = e^2\tilde q^2 \int \tilde T_h \wedge * \tilde T_h +
\dots \ ,
\label{r}
\end{equation}
which represents a generalized {\it Wilson `hypervolume law'}. This
shows
that the Julia-Toulouse mechanism describes the transition to a
{\it confinement phase} for the $(h-2)$-branes.

At this point we can ask what is the strong-weak coupling dual
formulation of ({\ref p}). This turns out to be
\begin{equation}
\tilde S_{d-h-1} = \int {e^2\over 4} d\tilde \phi _{d-h} \wedge *
d\tilde \phi _{d-h} + {\Lambda ^2\over 4} \left( \tilde \phi _{d-h}
-d\tilde \omega _{d-h-1} \right) \wedge * \left( \tilde \phi _{d-h}
-d\tilde \omega _{d-h-1} \right) \ ,
\label{s}
\end{equation}
where $d\phi _{h-1} = *d\tilde \phi _{d-h}$ and $d\omega _h =
*d\tilde \omega _{d-h-1}$. In this formulation we recognize a
generalized
version of the Higgs-St\"uckelberg mechanism: the original field
$\tilde \phi _{d-h}$ `eats' the new field $\tilde \omega _{d-h-1}$
due to the condensation of topological defects and acquires thereby a
mass $m=\Lambda /e$. We have thus shown that the Julia-Toulouse
mechanism
is the strong-weak coupling dual of a generalized Higgs-St\"uckelberg
mechanism. Correspondingly, the confinement phase for $\phi _{h-1}$
is
equivalent to a {\it Higgs phase} for its dual $\tilde \phi _{d-h}$.
This provides an explicit realization of the old ideas of 't Hooft
and
Mandelstam, along with a generalization to generic antisymmetric
tensor
field theories. It also provides a clue to non-perturbative
realizations
of strong-weak coupling dualities and shows that various
antisymmetric tensor field theories are connected by the condensation
of $p$-branes.$^{13}$

Up to now we have considered only the condensation of
$(d-h-1)$-branes.
Naturally, corresponding results are obtained due to the condensation
of
the dual $(h-2)$-branes. In this case one obtains a Higgs phase for
$\phi _{h-1}$, or, equivalently, a confinement phase for the dual
$\tilde \phi _{d-h}$.

The most interesting examples of this non-perturbative phases are
obtained for $h=(d+1)/2$. In these cases both types of topological
defects
have the same dimension and the Higgs and confinement phases are
described by tensors of the same rank. We thus expect a phase diagram
which is symmetric around the self-dual point $e^2=2$. In $D=2,4$
this
is triggered by instantons and monopoles, respectively, and is
realized
as follows: there is a Higgs phase for $\phi _{h-1}$ at weak coupling
and
a confinement phase for $\phi _{h-1}$ at strong coupling with the
possibility of an intermediate self-dual Coulomb (massless) phase for
$p$ in (\ref{o}) larger than a critical value.$^9$
The next interesting case regards string condensation in $D=6$. There
is
evidence that the same structure is realized.$^7$

To conclude this brief review we shall discuss two concrete examples
which are relevant for two different types of string theories. The
first
example is {\it compact QED}$^6$ in $(3+1)$ dimensions, with
(Euclidean) action
\begin{equation}
S= \int d^4x \ {1\over 4e^2} F_{\mu \nu }F_{\mu \nu } \ ,\qquad
\qquad
F_{\mu \nu }\equiv \partial _{\mu }A_{\nu }-\partial _{\nu }A_{\mu }
\ .
\label{t}
\end{equation}
This is a case for which $h=(d+1)/2$: both types of topological
defects are
$0$-branes, i.e. point particles. The topological defects of
(\ref{t})
are magnetic monopoles, while their dual topological defects are
electric charges, which have the standard Noether coupling to $A_{\mu
}$.

The condensation of electric charges leads to the familiar Higgs
phase
with effective low-energy action
\begin{equation}
S_H=\int d^4x \ {1\over 4e^2} F_{\mu \nu }F_{\mu \nu } + {\tilde
\Lambda ^2
\over 2} A_{\mu }A_{\mu } \ ,
\label{u}
\end{equation}
describing a massive vector of mass $m_H=e\tilde \Lambda $. In this
phase
the magnetic monopoles are confined, which is tantamount to the
Meissner
effect. The condensation of magnetic monopoles, instead, leads to a
confinement phase with effective low-energy action given by
\begin{equation}
S_C=\int d^4x \ {1\over 12\Lambda ^2} \partial_{[\mu }B_{\nu \alpha
]}
\partial _{[\mu }B_{\nu \alpha ]} + {1\over 4e^2} B_{\mu \nu }
B_{\mu \nu } \ ,
\label{v}
\end{equation}
describing a massive two-index tensor of mass $m_C= \Lambda /e$. In
this
phase electric charges are confined.

Actually the induced action
\begin{equation}
{\rm e}^{-S_{\rm ind}\left( \Sigma \right) } = {1\over Z\left( B_{\mu
\nu }
\right) } \ \int {\cal D}B_{\mu \nu } \ {\rm e}^{-S_C\left( B_{\mu
\nu }
\right) + i \int _{\Sigma } d\sigma _{\mu \nu } B_{\mu \nu } }
\label{w}
\end{equation}
defines the (low-energy) action for the (Abelian) {\it confining
string} with
world-sheet $\Sigma $, a result recently pointed out also by
Polyakov.$^{14}$

The second example we shall discuss is the puzzle of the axion mass
in 4D string models. In its simplest formulation, the axion is a
pseudoscalar with a `Noether' coupling to the QCD instantons:$^{15}$
\begin{equation}
S=\int d^4x \ {1\over 2} \partial _{\mu }a \partial _{\mu }a
+i a \ {1\over 16\pi ^2 f} {\rm Tr} F_{\mu \nu }{F_{\mu \nu }}^* \ .
\label{y}
\end{equation}
Instantons are always in a plasma phase for $D>2$. As is well
known$^{15}$
these instantons generate a potential $V(a)$ for the axion field. By
considering only small instantons this is given by $V(a)= \Lambda
_{QCD}^4
\left( 1-{\rm cos}\left( a/f \right) \right) $ which implies an axion
mass
$m_a= \Lambda _{QCD}^2/f$. In our terminology this would be a `Higgs
phase' for the axion.

In 4D string models, however, the axion appears in its dual
formulation,
given in terms of the Kalb-Ramond$^2$ two-index tensor $B_{\mu \nu
}$:
\begin{equation}
\tilde S=\int d^4x\ {1\over f^2} \left( \partial _{[\mu }B_{\nu
\alpha ]}
-K_{\mu \nu \alpha } \right) \left( \partial _{[\mu } B_{\nu \alpha
]}
-K_{\mu \nu \alpha } \right) \ ,
\label{yz}
\end{equation}
where $K_{\mu \nu \alpha}$ is the Chern-Simons three-form defined by
\begin{equation}
\epsilon_{\mu \nu \alpha \beta}\partial _{\mu }K_{\nu \alpha \beta} =
-{1\over 16 \pi ^2} {\rm Tr} F_{\mu \nu}{F_{\mu \nu }}^* \ .
\label{yw}
\end{equation}

Up until recently, the origin and description of the axion mass in
this
formulation were a puzzle. In our formalism, however, the solution of
this puzzle is very simple. Indeed, in the dual formulation small
instantons play the role of the topological defects of a compact
$B_{\mu \nu }$ field, the so-called axionic instantons.$^{6,16}$
Since these are always in a plasma phase we obtain the low-energy
effective action
\begin{equation}
\tilde S_{-1}= \int d^4x \ {1\over 4\Lambda _{QCD}^4} \partial _{[\mu
}
H_{\nu \alpha \beta ]} \partial _{[ \mu }H_{\nu \alpha \beta ]} +
{1\over f^2} H_{\mu \nu \alpha }H_{\mu \nu \alpha} \ ,
\label{z}
\end{equation}
formulated in terms of a {\it massive three-index tensor} of mass
$m_H=\Lambda _{QCD}^2/f$. This corresponds thus to a {\it string
confinement phase}. That in 4D strings are confined and turn to
membranes
has been, to our knowledge,
first proposed by Polyakov$^{17}$ and subsequently rediscovered by
S.-J. Rey.$^{16}$ Our result provides an explicit realization of this
mechanism and the low-energy effective action
for this confinement phase. Actually, with a construction analogous
to
(\ref{w}), the induced action obtained by coupling (\ref{z}) to the
world-volume of the membrane defines the action for the
`confining membrane'.

This example is particularly important since the result (\ref{z}) can
be obtained (it was actually first obtained this way) from the
supersymmetric version of (\ref{yz}) due to compactification of the
heterotic string.$^{18}$ In this case instantons drive gaugino
condensation
which implies the appearance of a massive three-index tensor in the
spectrum. This result can be considered as a supersymmetric version
of
the confining membrane. Moreover, it suggests the existence of a
generic
supersymmetric version of the Julia-Toulouse mechanism, which would
describe the transition to non-perturbative phases of string
theories.
Since the dilaton is always in the same supermultiplet with the
antisymmetric tensor, either it also acquires a mass or supersymmetry
is
broken in these phases. At least one of the outstanding problems of
string theory could be solved this way.

Let us finish with a description of the 2D scalar field $X$ living
on a circle of radius $R$, relevant for toroidal string
compactifications. Since $h=(d+1)/2=1$, the corresponding topological
defects are instantons and dual instantons. The action in the Coulomb
phase
is $4\pi\alpha'\, S=\int d^2 z\partial_\mu X\partial^\mu X+\cdots$
with periodicity $X\equiv X+2\pi R$. The dual action is identical in
terms
of the dual coordinate $\tilde X$ with periodicity $\tilde X\equiv
\tilde X+2\pi p \alpha'/R$, with $p$ the Dirac quantum
of equation (13). Following our prescription, the confinement phase
is described
by a massive vector dual to a massive scalar
$4\pi\alpha'\, S_C=\int d^2 z \left\{\partial_\mu\tilde
X\partial^\mu\tilde X-\Lambda^2R^2/\alpha'^2\, \tilde
X^2+\cdots\right\}$, with a similar expression for the Higgs phase
(exchanging $X\leftrightarrow \tilde X,
R\leftrightarrow p\alpha'/R, \Lambda\leftrightarrow\tilde{\Lambda}$)
and thus showing a Higgs/confinement duality. As mentioned above, the
phase diagram is symmetric around the selfdual point $R^2=p\alpha'$.
This was thoroughly studied in reference [9]  in terms of the `mass
formula'.
\begin{equation}
\alpha'\, M^2={p^2 n^2}\, \frac{\alpha'}{R^2}+m^2\,
\frac{R^2}{\alpha'}
+2\, \left(N_L+N_R-2\right).
\label{mass}
\end{equation}
If $X$ is a string coordinate, this is the standard mass formula
(for $p=1$)
with $n$ the quantized momentum, $m$ the winding number
and $N_{L,R}$ the left and right moving oscillator numbers.
It can be seen that the 2D
instantons interpolate among different windings and therefore
the dual instantons interpolate among different momenta.
Using the previous equation, Cardy and Rabinovici extracted the
phase diagram of this system [9]. A way to see this is to find
the points in $R$-space where tachyons appear. We can see that for
$n=N_{L,R}=0,\, m=\pm 1$ there is a tachyon for $R^2<4\alpha'$
whereas for $m=N_{L,R}=0,\, n=\pm 1$ there is a tachyon for
$R^2>p^2\, \alpha'/4$. For $p>4$ there is a range $4\alpha'<R^2<p^2\,
\alpha'/4$
without an extra tachyon and in that interval the (selfdual) Coulomb
phase is realized whereas the Higgs phase exists for larger values of
$R^2$,
dual to the confinement phase at smaller values of $R^2$. For $p<4$
there are only Higgs and confinement phases emanating at the selfdual
point. This is the case for modular invariant bosonic strings with
$p=1$.
This shows explicitly how $T$-duality is preserved in the presence
of nonperturbative effects once both instantons and dual instantons
are taken into account.

For perturbative supersymmetric strings there are no  tachyons in the
spectrum for any value of $R$ and then only the Coulomb phase is
realized. However, at
finite temperature, the situation is again interesting. In this case,
$X$ is the time coordinate and the temperature is the inverse period
$T=1/2\pi R$.
The mass formula is (for $p=1$) [19]:
\begin{equation}
\alpha'\, M^2={n^2}\, \frac{\alpha'}{R^2}+m^2\, \frac{R^2}{\alpha'}
+2\,  \left(N_L+N_R-\frac{3}{2}\right),
\label{massh}
\end{equation}
with the level matching condition: $mn=\left(N_R-N_L+1/2\right)$.
The difference with the purely bosonic case relies on the fact that
$n$ can be half-integer. Actually the states with $m=\pm 1, n=m/2,
N_{L,R}=0$ are tachyonic in the range
$1+1/\sqrt{2}>R/\sqrt{\alpha'}>1-1/\sqrt{2}$.
Since the tachyonic states carry both momentum and winding,
the corresponding instantons will be `dyonic instantons'.
Therefore there is a Coulomb phase at  temperatures  below
the Hagedorn temperature $T_H\equiv (\sqrt{2}-1)/\sqrt{2\alpha'}\pi$,
where the condensation of `mdyons' induce a phase transition.
This (selfdual) phase finishes at the dual Hagedorn temperature
$\tilde{T}_H\equiv (\sqrt{2}+1)/\sqrt{2\alpha'}\pi$ where the tachyon
disappears and  the dual Coulomb phase starts.

The physical picture we get is the following:
for temperatures $T<T_H$ the system is in a Coulomb phase described
in
terms of a worldsheet action with compact euclidean time defining
temperature.
At temperatures $T_H<T<\tilde{T}_H$ there is the `dyon condensate'
phase, described by a massive vector field dual to a `massive time
coordinate'.
Since standard euclidean time is no longer well defined in this
phase,
$T$ may no longer be the physical temperature.
Instead, an effective temperature $T_{eff}$ depending on
both $T$ and the mass parameter $\Lambda$ will have to
be defined up to the selfdual point. Beyond that point, the whole
picture is repeated but now in terms of the dual variables. This
might
solve the problem of how to make compatible temperature duality with
the fact that the partition function should be a monotonic
function of temperature [19].
Therefore, temperature duality may also be an exact symmetry
of string theory.  This could
 have very interesting cosmological implications.
It may also be interesting to find a supersymmetric generalization
of our approach to  compare with the results of ref. [20] where  a
target space formulation
was used.

In summary we have  seen how,  in our
approach, duality is always a good nonperturbative symmetry
once all the relevant topological defects are included.
The generalized Higgs and confinement phases we have described may
 actually be  present in nonperturbative string theory where
$p$-branes are playing a fundamental role. We may actually
be in a $p$-brane confining phase where the dilaton field is no
longer massless.


\nonumsection{Acknowledgements}
\noindent
C. A. Trugenberger is supported by a Profil 2 fellowship of
the Swiss National Science Foundation.

\nonumsection{References}

\end{document}